# TinyLIC-high efficiency lossy image compression method

Gaocheng Ma, Yinfeng Chai and Tianhao Jiang
Ming Lu, Tong Chen
*Nanjing University*
*{211180094,211180091,211180078}@smail.nju.edu.cn*
*{minglu,chentong}@nju.edu.cn*

**Abstract**: Image compression has been the subject of extensive research for several decades, resulting in the development of well-known standards such as JPEG, JPEG2000, and H.264/AVC. However, recent advancements in deep learning have led to the emergence of learned image compression methods that offer significant improvements in coding efficiency compared to traditional codecs. These learned compression techniques have shown noticeable gains and even outperformed traditional schemes.

## 1. Introduction:

Conventional image compression methods usually apply prediction, transformation, and quantization successively to achieve compression. On the other hand, learning image compression methods make use of deep neural networks to optimize the compression process, replacing handcrafted components with fully learned ones. However, these methods still face challenges in practical applications due to the high spatio-temporal complexity requirements.

In this paper, we have adopted a new approach to image compression using quality scaling factor (SF). The main objective is to encode images of various bit rates with a single model, thus reducing the complexity of the implementation. By embedding SFs into the compression network in this process, performance comparable to existing methods that require multiple bitrate-dependent models can be achieved. Inspired by the quantization scaling mechanism used in **H.264/AVC** and **JPEG**, the approach used exploits the concept of scaling factors in the compression framework. Applying the scaling factor to the underlying features of the image, this approach provides a practical and efficient solution for variable bitrate image compression.

We evaluated their method using a variety of compression network structures, image contents, and training loss functions. The results show that our method is general and efficient in practical applications. The proposed method reduces space consumption and computational overheads, making it an optimal solution for practical image compression applications.

In conclusion, we used a new method for image compression using quality scaling factor. The method provides a practical solution for encoding images of various bit rates using a single model, thereby reducing complexity and increasing efficiency.

## 2. Proposed Method

This section first overviews the proposed **TinyLIC**. More details are given subsequently for transform and entropy coding. For better comprehension, notations are given in *Table I*

### 2.1 Overview

Figure 2 illustrates the **TinyLIC** which follows the end-to-end **VAE** architecture [1] to construct and aggregate encoder-decoder pairs for rate-distortion optimized compact representation. Given an input image $x \in R^{H \times W \times 3}$, the analysis transform $g_\theta(\cdot)$ is firstly applied to extract the latent representation $z$, which is further processed to generate hyper features z through $h_\varphi(\cdot)$. A quantizer is used to discretize $z$ and $\tilde{z}$ for compact representation. A simple factorized entropy model is applied for the entropy coding of $z$ and $\tilde{z}$. The decomposed conditional model is used to predict the probability distribution where its mean μ and scale σ are predicted using decoded neighboring blocks [1], for instance, $p_\mu(\tilde{z}), \sigma(\tilde{z})$, following a pre-arranged order. The final $\tilde{x}$ is reconstructed using the same analysis transform $g_\theta(\cdot)$.

Therefore, the rate-distortion optimization of the **VAE** in Figure 2 can be extended from (1) as:

$$J = \mathbb{E}_{p_x}[-\log p_z(g_\theta(x)|z)] + \mathbb{E}_{p_x}[z \sim p_z(z|x)][-\log p_z(z)]$$

where $p_\chi$ is the distribution of the input source image, $p_\nu$, and $p_z$ are the probability distributions of respective y and z at the bottleneck layer for entropy coding.

Next, the **ICSA-based** content-adaptive transform and the **MCM-based** context modeling are detailed.

| Abbr. | Descriptions |
|---|---|
| ANIA | Adaptive Neighborhood Information Aggregation |
| ICSA | Integrated Convolution and Self-Attention |
| RNAB | Residual Neighborhood Attention Block |
| MCM | Multi-stage Context Model |
| GCP | Generalized Checkerboard Pattern |
| VAE | Variational Auto-Encoder |
| LIC | Lossy Image Coding |
| MAC | Multiply-Accumulate Operation |
| BD-rate | Bjøntegaard Delta Rate |
| PSNR | Peak Signal-to-Noise Ratio |
| MSE | Mean Square Error |
| MS-SSIM | Multi-Scale Structural Similarity |
| BPG | Better Portable Graphics |
| HEVC | High-Efficiency Video Coding |
| VVC | Versatile Video Coding |

Table 1  **Notations**

## 2.2 Content-Adaptive Transform via Stacked ICSAs
### 2.2.1 Convolutional Feature Embedding and Resampling:
To ensure the robustness of spatial consistency as well as early vision transformations from studies [7], [8], we perform the convolutional feature embedding to input the ICSA into the latent space as shown in *Figure 2*. Some tokenizations are enforced for subsequent ICSA unit processing.

With the RANB module for proceeding ICSA for hierarchical feature embedding, it is noteworthy that we also apply spatial resampling at the convolution layer in each ICSA. This resampling enlarges the receptive field to some extent, allowing for subtle downsampling of the spatial resolution with negligible information loss. For simplicity, we apply uniform sampling at each dimension with a stride of 2. In order to encode the spatial relationship position information [11], without the need for explicit position signals as reported in [7], [8]. Additionally, convolutional features from overlapping pixel patches used in [10], are applied for subsequent window-based self-attention computation to avoid block artifacts, and are beneficial to early visual processing and stable training as reported in [12], [13], and our simulations in Sec. V.A.

### 2.2.2 Window-based Self-Attention via RANB:

Despite the great successes of high-level vision tasks [7], [8], it is challenging to migrate the attention layer (e.g., self-attention) from high-level vision tasks to lower-level computation tasks (e.g., compression), due to the complexity of transferring the self-attention layer in Transformers to low-level computation of window-based input image size. Recently, the Neighborhood Attention Transformer (NAT) was proposed in [6] as an example to form the Residual Neighborhood Attention Block (RANB) for window-based self-attention computation. Other window-based self-attention mechanisms like Swin Transformer [9] can be adopted as well (see Sec. V.A). Multiple RANBs are often stacked and connected with a convolutional layer to form an ICSA unit as shown in Figure 2.

Figure 3 briefly sketches the processing flow of an RANB.
First, the feature embedding process (FE) transforms an input image or feature tensor at a size of $H_f \times W_f \times C$ to a dimension of $H_f / W_f \times C$ for processing.

Subsequently, the NAT consolidates neighborhood information by stacking layers that separately undertake neighborhood attention processing (NA), multi-layer perceptron processing (MLP), and layer normalization (LN).

Finally, the feature de-embedding layer (FU) maps the attention-weighted features back to the original resolution dimensions of $H \times W \times \mathbb{C}$.

Following standard practice, residual skip connections are used to bolster information aggregation and model training efficacy.
Feature aggregation within the NA layer is articulated as follows

$$NA(i,j) = softmax\left(\frac{\mathbf{Q}(i,j) K^T}{\sqrt{d}}\right) V(i,j)$$

Here, (i,j) indicates the element's coordinates, and $Q$, $K$, and $V$ signify the features transformed by linear operations corresponding to query, key, and value, respectively. $B_{ij}$ denotes the positional bias learned. The **MLP** layer is coupled with an activated **GELU** layer. As depicted, $Q(i,j)$, $K$ and $V$ are immediately utilized in computations to weight local information, which in turn reflects the content attributes of any given input.

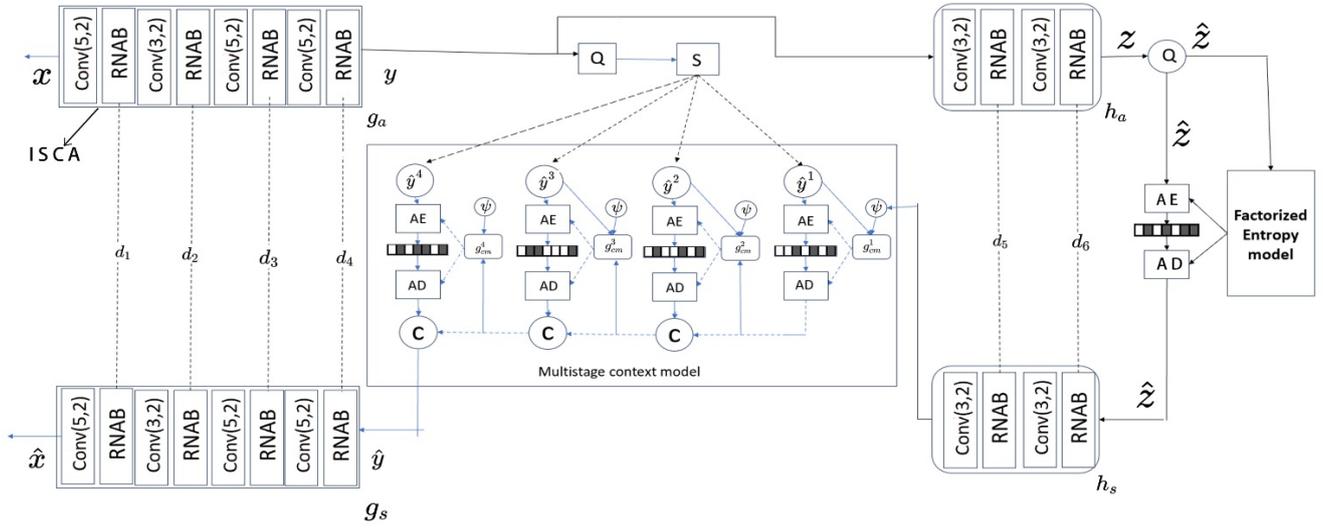

Fig. 2. **TinyLIC**. The prominent VAE structure uses a master encoder and a hyperencoder pair. Four and two paired ICSA units are used in the main encoder and the hyper encoder. Each ICSA consists of a convolutional layer for feature embedding and spatial resamples, and multiple RNABs aggregate neighborhood information via attention-based weighted adaptive aggregation. Di, I = 1,2..., 6 is the amount of RNAB used in stage i. Convolution Conv(k,s) and its transposed version TConv(k,s) apply to nuclei of size k × k, step s, k = 3 or 5,s = 2. MCM makes full use of the channel and spatial elements of the superpriori ψ and non-uniform grouping to obtain better probability estimates, and uses a simple decomposition entropy model to encode the superpriori. S and C represent slicing and concatenation of tensors; Q Adopt uniform quantization; AE and AD stand for arithmetic encoding and arithmetic decoding respectively

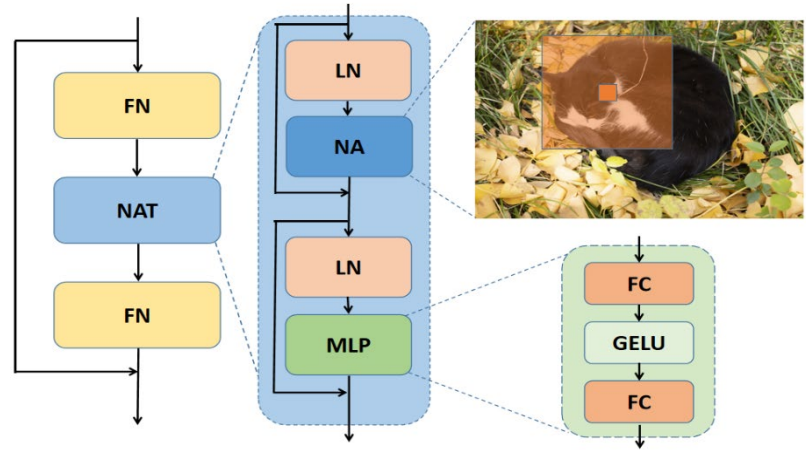

Fig. 3. **Residual Neighborhood Attention Block**. FE and FU represent the feature embedding layer and the feature solution embedding layer respectively. NAT is a neighborhood attention converter consisting of a layer normalization (LN) layer, a neighborhood attention (NA) layer and a multi-layer perceptron (MLP) layer. Two fully connected (FC) layers are interwoven with a GELU activated layer to form the MLP layer. Apply the remaining skip connections.

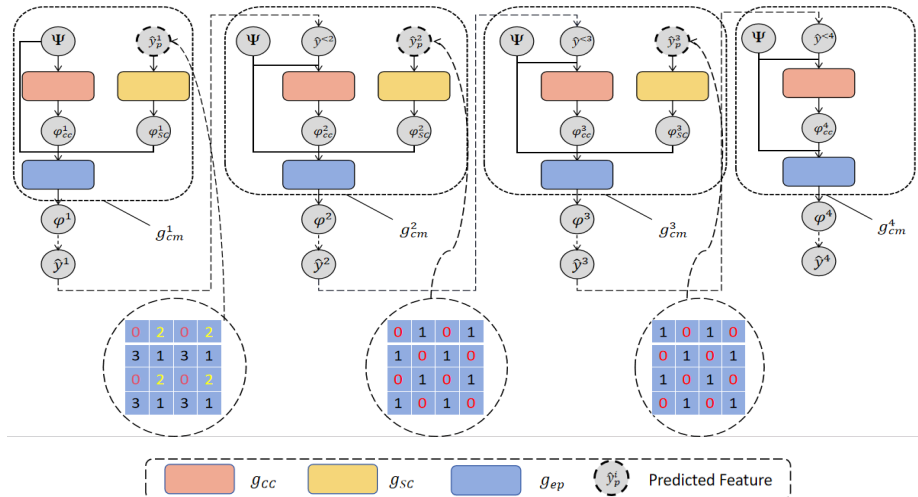

Fig. 4. **Multistage Context Model.** The channel condition model and the space condition model were used for contextual prediction. The encoded feature segment has been preprocessed contextually.

## 2.3 Multistage Context Model

Utilizing hyperparameters that jointly control spatial-channel neighbors for context modeling greatly enhances the rate-distortion performance [1][2][4]. However, the encoding runtime's excessive computational load often leads to the sequential processing of aggressive neighbor features by contextualizing each element in isolation.

Given the strong correlation among spatial-channel neighbors, including the sequential model within our autoregressive model can enhance accuracy. We execute the Multi-channel Context Model (MCM) in parallel, aiming for precise probability predictions for the spatial-channel neighbor features. As illustrated in Figure 4, the Four-Stage processing pipeline of the proposed MCM meticulously segregates feature tensors by the channel dimension for detailed processing.

For instance, the feature tensor $\hat{y}_i$ is divided into four groups along the channel dimension. During the learning stage, these groups are scaled with varying channels to boost generalization capacity. A checkerboard strategy is employed for spatial partitioning to upscale the dimensionality of the feature tensors to $H \times W \times C$ for subsequent stages.

In the first stage, we refine the process using Grid Conditioned Probability (GCP) methods, utilizing a 4-step GCP for finer spatial utilization. At this phase, fine-grained $1 \times 1$ convolutions are used for deriving entropy parameters, i.e., mean μ and scale σ. Then tensor concatenation is carried out for processing the second stage.

During the second stage, the process involves $\sigma^2$, where $\sigma^2$ is computed using $g_C^2((y_n^1, \hat{y}_{\phi(n)}^1))$ with $g_C^2$ being the spatial aggregation processing function for the second stage, taking $\hat{y}_{\phi(n)}^1$ as inputs for different spatial-channel neighbors.

Similarly, the third and fourth stages proceed with analogous methodologies to enhance the overall model performance.

## 3. Discussion

It also evidences that the VAE architecture with nonlinear transform and entropy context conditioned on joint hyperpriors and spatial-channel neighbors which was first proposed in Minnen'18 is a well-generalized solution regardless of the different techniques used in its modular components like ReLU or GDN-based activation, simple convolution or ICSA, etc [1], [2], [4], [5]. When comparing to coding efficiency offered by Minnen'18, almost 10 absolute percentage points improvement is reported against the same BPG anchor. The attention mechanism further reveals its outstanding effectiveness in adaptively weighing and aggregating highly correlated information from the results of Cheng'20 [2] and our TinyLIC. Compared with Cheng'20 [2], our method not only extends the attention embedding to all stages (but not just the bottleneck layer) in main and hypercoders but also replaces the convolution-based attention computation with the self-attention to flexibly characterize any dynamic input.

The VVC Intra is expected to succeed its predecessor HEVC Intra (BPG) because of its outstanding performance [3]. We then switch the anchor from BPG to VVC Intra to derive coding gains of the proposed TinyLIC, for which ≈3% BD-rate gains are captured on average across all three datasets.

## 4. Conclusion

A novel learned image coding method - TinyLIC was developed in this work, presenting the superior compression performance and high-throughput computation Joint high-performance compression and high throughput computation of the proposed TinyLIC comes from the intelligent use of adaptive neighborhood information aggregation. To this end, we integrate the convolution and self-attention to form the content-adaptive transform by which we can dynamically characterize and embed the neighborhood information conditioned on the input content; we further propose the multistage context model using local spatial channel neighbors in a managed order for entropy coding to unknit the autoregressive dependency for parallel processing while still retaining the efficiency as the autoregressive model.

# 5.References

[1] D. Minnen, J. Balle, and G. Toderici, "Joint autoregressive and hier-´ archical priors for learned image compression," in *Advances in Neural Information Processing Systems*, 2018, pp. 10 794–10 803

[2] Z. Cheng, H. Sun, M. Takeuchi, and J. Katto, "Learned image compression with discretized gaussian mixture likelihoods and attention modules," in *Proceedings of the IEEE/CVF Conference on Computer Vision and Pattern Recognition*, 2020, pp. 7939–7948.

[3] J. Pfaff, A. Filippov, S. Liu, X. Zhao, J. Chen, S. De-Luxan-Hern´ andez, ´ T. Wiegand, V. Rufitskiy, A. K. Ramasubramonian, and G. Van der Auwera, "Intra prediction and mode coding in vvc," *IEEE Transactions on Circuits and Systems for Video Technology*, vol. 31, no. 10, pp. 3834– 3847, 2021.

[4] T. Chen, H. Liu, Z. Ma, Q. Shen, X. Cao, and Y. Wang, "End-toend learnt image compression via non-local attention optimization and improved context modeling," *IEEE Transactions on Image Processing*, vol. 30, pp. 3179–3191, 2021.

[5] M. Lu, P. Guo, H. Shi, C. Cao, and Z. Ma, "Transformer-based image compression," in *IEEE Data Compression Conference*, 2022

[6] A. Hassani, S. Walton, J. Li, S. Li, and H. Shi, "Neighborhood attention transformer," *arXiv preprint arXiv*:2204.07143, 2022

[7] N. Carion, F. Massa, G. Synnaeve, N. Usunier,A. Kirillov, and S. Zagoruyko, "End-to-end object detection with transformers," in *European Conference on Computer Vision. Springer*, 2020, pp. 213– 229.

[8] A. Dosovitskiy, L. Beyer, A. Kolesnikov, D. Weissenborn, X. Zhai, T. Unterthiner, M. Dehghani, M. Minderer, G. Heigold, S. Gelly et al., "An image is worth 16x16 words: Transformers for image recognition at scale," *arXiv preprint arXiv*:2010.11929, 2020.

[9] Z. Liu, Y. Lin, Y. Cao, H. Hu, Y. Wei, Z. Zhang, S. Lin, and B. Guo, "Swin transformer: Hierarchical vision transformer using shifted windows," *International Conference on Computer Vision* (ICCV), 2021.

[10] Y. Zhu, Y. Yang, and T. Cohen, "Transformer-based transform coding," in *International Conference on Learning Representations*, 2022.

[11] M. A. Islam, S. Jia, and N. D. Bruce, "How much position information do convolutional neural networks encode?" in *International Conference on Learning Representations*, 2019.

[12] A. Gulati, J. Qin, C.-C. Chiu, N. Parmar, Y. Zhang, J. Yu, W. Han, S. Wang, Z. Zhang, Y. Wu et al., "Conformer: Convolution-augmented transformer for speech recognition," *arXiv preprint arXiv*:2005.08100, 2020.

[13] T. Xiao, M. Singh, E. Mintun, T. Darrell, P. Dollar, and R. Gir- ´ shick, "Early convolutions help transformers see better," *arXiv preprint arXiv*:2106.14881, 2021.